\documentclass{amia}
\usepackage{graphicx}
\usepackage[labelfont=bf]{caption}
\usepackage[superscript,nomove]{cite}
\usepackage{color}
\usepackage{xcolor}

\usepackage{soul}

\begin{document}

\title{Addressing the Need for Remote Patient Monitoring Applications in Appalachian Areas}

\author{
    Alyssa Donawa, %
    Corey E. Baker
}

\institutes{
    University of Kentucky, Lexington, KY, USA
}

\maketitle

\noindent{\bf Abstract}
\textit{
There is a need to address the urban-rural disparities in healthcare regarding equal access and quality of care.
Due to higher rates of chronic disease, reduced access to providers, %
and a continuous decline in rural hospitals,
it is imperative that Appalachian cancer patients adopt the use of health information technology (HIT). %
The \textit{NCCN Distress Thermometer and Problem List} (DT) is under-utilized, not patient-centered, does not consider provider needs, and is outdated in the current digital landscape.
Digitizing patient distress screening 
poses advantages, such as allowing for more frequent screenings, removing geographical barriers, and rural patient autonomy.
In this paper, we discuss how knowledge gained from patient-centered design led to the underpinnings of developing a rural remote patient monitoring app that provides delightful and insightful experiences to users.
}

\section*{Introduction}
According to the National Cancer Institute,
distress is an emotional, social, spiritual, or physical pain or suffering that may cause a person to feel sad, afraid, depressed, anxious, or lonely. 
Distress is highly prevalent in cancer patients regardless of disease-stage or modality.~\cite{albrecht2012management}
In the case of cancer patients, untreated distress has been shown to lead to greater pain, reduced physical function, increased medical costs, and longer stays in the hospital. 
The National Comprehensive Cancer Network (NCCN) designed %
the \textit{NCCN Distress Thermometer and Problem List} (DT) %
to be used as a screening tool for recognizing distress in cancer patients%
; and has since been shown to accurately indicate distress.~\cite{zebrack2017practice} %
The DT was designed to improve patient care, which in turn would improve a patient's quality of life. 
Furthermore, studies have shown that routine distress screening is able to improve health outcomes including %
morbidity and mortality.~\cite{basch2017overall} %
A greater issue emerges when considering rural communities in the United States that are often medically underserved and %
medically disadvantaged, such as Appalachian Kentucky.~\cite{hesse2020barn}
Rural communities also %
commonly have higher rates of chronic disease, reduced access to providers, %
and continue to experience a decline in rural hospitals.~\cite{saslow_2019,national2017health} %
The aforementioned disparities raise the urgency %
for rural communities to adopt the use of health information technologies (HIT). %
However, rural communities face geographical and financial challenges that result in limited or nonexistent access to broadband connectivity. %
This ``digital divide" limits the ability for rural communities to benefit from HIT.~\cite{max2020augmenting,hesse2020barn}
In addition,
rural communities also have lower levels of overall technology adoption.~\cite{hesse2020barn, mccomsey2020experiencing} %
\textbf{In this paper, we discuss how knowledge gained from patient-centered design led to the underpinnings of developing a rural remote patient monitoring app that provides delightful and insightful experiences to users.}

\section*{Methodology}
In order to better understand how distress manifests for cancer patients with ties to rural Kentucky,
four participatory design workshops were held between February and October of %
2019 %
to gather input from local stakeholders.~\cite{hesse2020barn} %
These co-design workshops gave community participants a way to directly contribute to the brainstorming, design, and low-fidelity prototyping of distress monitoring tools that would positively impact the cancer experience. %
We were able to bring together various stakeholders from the %
Appalachian Kentucky and Lexington communities which included medical professionals, social workers, technologists, students, current and previous cancer patients, and researchers. %
Knowledge was gained regarding what stakeholders would like to see in HIT and distress screening tools.
Understanding was also gained on the underlying holistic needs of Appalachian cancer patients; an example being that patients may feel embarrassed to discuss their distress symptoms out loud, which relates to the culture of self-sufficiency that is prevalent in Appalachian culture.~\cite{mccomsey2020experiencing}  %
Following the workshops, participants were invited to provide feedback on the DT and the digital translation of the DT utilizing the System Usability Scale (SUS).
 Community members at the Markey Cancer Center and attendees of the Markey Cancer Center Affiliate Network (MCCAN) 2019 Cancer Care Conference were also invited to score the distress screening tools.
 The sample of people who scored the paper DT was $n=44$, with 8 identifying as patients, 10 as caregivers, 11 as providers, and 15 as other.
 The sample of people who scored the digital DT was $n=34$, with 
7 identifying as patients, 10 as caregivers, 8 as providers, and 9 as other.
All users reported that they preferred the digitized distress screening tools over paper tools; supporting our decision to convert cancer patient distress screening to a digital format for target users.
\section*{Assuage - An App for Rural Health Care}
Assuage is a HIPAA compliant mobile iOS %
application by researchers at the University of Kentucky (UK). \footnote{Network Reconnaissance Lab, University of Kentucky - https://www.cs.uky.edu/~baker/research/} 
An aim of Assuage is to enhance the process of distress monitoring in rural cancer patients with more frequent screening.
Cancer patients at UK's Markey Cancer Center complete the DT approximately every six weeks.
By reducing the time between screenings,
providers and researchers can better understand a patient's overall distress, causes of distress, and track symptoms between visits.
Assuage also seeks to facilitate patient provider communication.
In the context of Assuage, all DT components are referred to as ``surveys''.
Assuage offers %
the ability to choose from four different user interfaces (UIs) in order to complete the %
routine distress assessment.
The decision to offer multiple UIs was made with knowledge that Appalachians have not heavily adopted HIT, but are also not completely removed from modern everyday technologies, like cellphones.
This takes a different approach than related work that seeks to overcome challenges with rural cancer patients and information access,~\cite{jacobs2018mypath} or daily life management following cancer diagnosis.~\cite{khurram2020patient}
In order to ensure usability and routine completion of the distress assessment, %
multiple UIs are offered to gain understanding of patient preferences.
While Assuage is still %
undergoing final iterations before the initial pilot, several of the current features include: %
\\
{\textbf{Health App Integration.}}
Assuage leverages Apple's HealthKit, CareKit, and ResearchKit\footnote{Apple, ResearchKit and CareKit - https://www.researchandcare.org} to provide an engaging experience for both Patients and Doctors. Assuage is also able to collect information from any Bluetooth based sensor.
The first time a user logs in to Assuage, they are prompted with the option to allow Assuage to gain access to data from the Health app.
Following this step, users are launched into the Assuage app.
Note that this will happen on the first installation and sign in of the Assuage app.
Afterwards, users can %
manually change their preferences through their iPhone's general settings.
\\
\textbf{Multiple UIs.}
In order to solicit feedback on UI preferences, Assuage offers patient users the ability to choose from four different UIs in order to complete their routine distress assessment.
The UIs differ by the way the surveys are displayed and navigated.
(1) The \textit{NCCN Advanced} UI
implements a modularized view of the DT components.
Users can select \textit{cards} corresponding to surveys; allowing for the most fluid navigation %
between sections.
(2) The \textit{NCCN Standard} UI guides patients sequentially through the surveys.
Navigation is limited to \textit{next} and \textit{back} buttons.
(3) The \textit{NCCN Checklist} UI presents patients with the option to navigate sequentially through the surveys, similar to the previous UI, or by selecting buttons with the associated survey labels.
The latter allows for customized and more direct navigation of surveys.
The label/button associated with the current survey will be highlighted. %
(4) The \textit{NCCN Paper} %
UI offers the smoothest transition for patients who prefer the standard paper DT.
Patients take a photo of their manually completed paper DT using their device; and upload it in Assuage.
\\
\textbf{Wireless.} %
Assuage will be functional in a fully connected and intermittently connected network;
addressing the barrier of limited broadband connectivity in Appalachian Kentucky by using device-to-device (D2D) communication %
that utilizes the mobility of rural residents to maximize data delivery.~\cite{max2020augmenting}
Note that Assuage is not intended for use in medical related emergencies requiring immediate attention, but for patient monitoring, feedback, and updates.
\section*{Future Work}
While the primary goal of Assuage is to improve routine distress screening for rural cancer patients, we hope that with continued use rural patients will more readily adopt the use of other HIT.
Future iterations of Assuage will be designed to have interfaces for non-patient users; the first priority being a care provider/doctor interface.
A feasibility study of Assuage with Appalachian patients is also necessary before moving forward with feature refinement.
We hope to leverage this desire for self-sufficiency coupled with the desire for improved health outcomes %
to incentivize the sustained use of Assuage. %
 \bibliographystyle{unsrt}
 \setlength\itemsep{-0.1em}
\bibliography{Ref}

\begin{thebibliography}{10}

\bibitem{albrecht2012management}
Tara~A Albrecht and Margaret Rosenzweig.
\newblock Management of cancer related distress in patients with a
  hematological malignancy.
\newblock {\em Journal of hospice and palliative nursing: JHPN: the official
  journal of the Hospice and Palliative Nurses Association}, 14(7):462, 2012.

\bibitem{zebrack2017practice}
Brad Zebrack, Karen Kayser, Deborah Bybee, Lynne Padgett, Laura Sundstrom, Chad
  Jobin, and Julianne Oktay.
\newblock A practice-based evaluation of distress screening protocol adherence
  and medical service utilization.
\newblock {\em Journal of the National Comprehensive Cancer Network},
  15(7):903--912, 2017.

\bibitem{basch2017overall}
Ethan Basch, Allison~M Deal, Amylou~C Dueck, Howard~I Scher, Mark~G Kris,
  Clifford Hudis, and Deborah Schrag.
\newblock Overall survival results of a trial assessing patient-reported
  outcomes for symptom monitoring during routine cancer treatment.
\newblock {\em Jama}, 318(2):197--198, 2017.

\bibitem{hesse2020barn}
Bradford~W Hesse, David Ahern, Michele Ellison, Eliah Aronoff-Spencer, Robin~C
  Vanderpool, Karen Onyeije, Michael~C Gibbons, Timothy~W Mullett, Ming-Yuan
  Chih, Victoria Attencio, et~al.
\newblock Barn-raising on the digital frontier: The launch collaborative.
\newblock {\em Journal of Appalachian Health}, 2(1):6--20, 2020.

\bibitem{saslow_2019}
Eli Saslow.
\newblock `urgent needs from head to toe': This clinic had two days to fix a
  lifetime of needs, Jun 2019.

\bibitem{national2017health}
National~Center for Health~Statistics.
\newblock {\em Health, United States, 2016, with chartbook on Long-term trends
  in health}.
\newblock Number 2017. Government Printing Office, 2017.

\bibitem{max2020augmenting}
Esther Max-Onakpoya, Oluwashina Madamori, Faren Grant, Robin Vanderpool,
  Ming-Yuan Chih, David~K Ahern, Eliah Aronoll-Spencer, and Corey~E Baker.
\newblock Augmenting cloud connectivity with opportunistic networks for rural
  remote patient monitoring.
\newblock In {\em 2020 International Conference on Computing, Networking and
  Communications (ICNC)}, pages 920--926. IEEE, 2020.

\bibitem{mccomsey2020experiencing}
Melanie McComsey, David~K Ahern, Robin~C Vanderpool, Timothy~W Mullett,
  Ming-Yuan Chih, Meghan Johnson, Michele Ellison, Karen Onyeije, Bradford~W
  Hesse, and Eliah Aronoff-Spencer.
\newblock Experiencing cancer in appalachian kentucky.
\newblock {\em Journal of Appalachian Health}, 2(3), 2020.

\bibitem{jacobs2018mypath}
Maia Jacobs, Jeremy Johnson, and Elizabeth~D Mynatt.
\newblock Mypath: Investigating breast cancer patients' use of personalized
  health information.
\newblock {\em Proceedings of the ACM on Human-Computer Interaction},
  2(CSCW):1--21, 2018.

\bibitem{khurram2020patient}
Sadaf Khurram and Khurram Sardar.
\newblock Patient-centric mobile app solution.
\newblock In {\em Proceedings of the Australasian Computer Science Week
  Multiconference}, pages 1--4, 2020.

\end{thebibliography}

\end{document}